# HIGH TRANSVERSE MOMENTUM INCLUSIVE NEUTRAL PION PRODUCTION IN d+Au COLLISIONS AT RHIC

O. GREBENYUK[1], A. MISCHKE[1] and A. STOLPOVSKY[2]
for the STAR collaboration

[1]NIKHEF Amsterdam/Utrecht University, The Netherlands, oleksandr.grebenyuk@nikhef.nl
[2]Wayne State University, USA, alexst@physics.wayne.edu

*Abstract.* Preliminary results are presented on high $p_T$ inclusive neutral pion production in d+Au collisions at $\sqrt{s_{NN}} = 200$ GeV in the pseudo-rapidity range $0 < \eta < 1$. Photons from the decay $\pi^0 \rightarrow \gamma\gamma$ are detected in the Barrel Electromagnetic Calorimeter of the STAR experiment at RHIC. The analysis procedure is described in detail and the results are found to be in good agreement with earlier STAR findings and with next-to-leading order perturbative QCD calculations.

## 1.  INTRODUCTION

The high centre-of-mass energy of $\sqrt{s_{NN}} = 200$ GeV available at the Relativistic Heavy Ion Collider (RHIC) at Brookhaven opens up the hard scattering regime in heavy-ion collisions. Hard particles at large $p_T$ originate from the early stage of the interaction and therefore probe the medium produced in these collisions.

Among the most important findings at RHIC to date are the suppression of high $p_T$ particle yields in central Au+Au collisions compared to $N_{bin}$ scaled proton-proton interactions and the disappearance of jet-like correlations opposite to trigger hadrons [1, 2, 3]. No suppression effects were seen in d+Au collisions which lend support to the ansatz that suppression is due to parton energy loss by induced gluon radiation in an extremely dense medium. For a recent review see [4]. To quantitatively understand the medium induced modification of hadron production, precise measurements of identified hadrons at high $p_T$ in d+Au are required. The STAR Barrel Electromagnetic Calorimeter (BEMC) allows high transverse momentum measurements of $\pi^0$, $\eta$ and direct photons and may also contribute to the identification of $\rho$ mesons in the $\pi\pi$ decay channel.

## 2. THE STAR ELECTROMAGNETIC CALORIMETER

STAR uses the Barrel Electromagnetic Calorimeter (BEMC) [5] to trigger on high $p_T$ photons and electrons for the study of high $p_T$ particle production (like $\pi^0$ and $\eta$) and rare probes (jets, direct photons and heavy quarks). The BEMC surrounds the TPC tracker at a radius of 220 cm and covers full azimuth and a pseudorapidity range of $-1 < \eta < 1$. During the d+Au run in 2003 only half of the BEMC was operational covering $0 < \eta < 1$ (in the direction of the deuterium beam).

The BEMC is a lead-scintillator sampling calorimeter with a depth of 21 radiation lengths made of 4800 towers with granularity $(\Delta\eta, \Delta\varphi) = (0.05, 0.05)$. Two layers of gaseous shower maximum detectors (SMD, see Fig. 1), located approximately at a depth of $5X_0$ in the calorimeter module, measure the electromagnetic shower position and shape with high resolution $(\Delta\eta, \Delta\varphi) = (0.006, 0.006)$. A pre-shower sampling of the signal in the first two layers contributes to the hadron/photon discrimination.

The energy calibration of individual calorimeter towers is based on the minimum ionizing particle (MIP) response to charged hadrons measured in the TPC [6], using the expected MIP response from test-beams [7]. The overall energy scale is determined from the response to high-$p_T$ electrons that are identified in the TPC. The intrinsic energy resolution due to sampling fluctuations is $\Delta E/E \approx 16\%/\sqrt{E}$. The dynamic range for photon detection is approximately 1 – 25 GeV/c.

The minimum bias interaction trigger required the detection of at least one neutron by the Zero Degree Calorimeter in the Au beam direction. The trigger acceptance is $(95\pm3)\%$ of the d+Au hadronic cross section. To enhance the yields at high $p_T$, two high tower triggers (HT1 and HT2) were used which required the energy deposition in at least one calorimeter tower to be above an effective $p_T$ threshold of 2.5 and 4.5 GeV/c, respectively. For d+Au events the tower occupancy is a few percent and the HT trigger efficiency is almost 100%.

## 3. ANALYSIS AND RESULTS

In this paper we present progress in the systematic studies and technical improvements of the STAR $\pi^0$ analysis. A first report of this analysis can be found in [8]. For the analysis 10M d+Au events were used after event quality cuts. The extensive tower-by-tower QA was performed to identify malfunctioning towers. Neutral pions were reconstructed in the decay channel $\pi^0 \to \gamma\gamma$ by calculating the invariant mass of photon-like calorimeter hit pairs which does not have pointing TPC tracks (Fig. 1).

A cut on the two-particle energy asymmetry $|E_1 - E_2|/(E_1 + E_2) < 0.5$ was imposed to increase the signal to background ratio and a SMD signal was required for both particles to improve the spatial resolution.

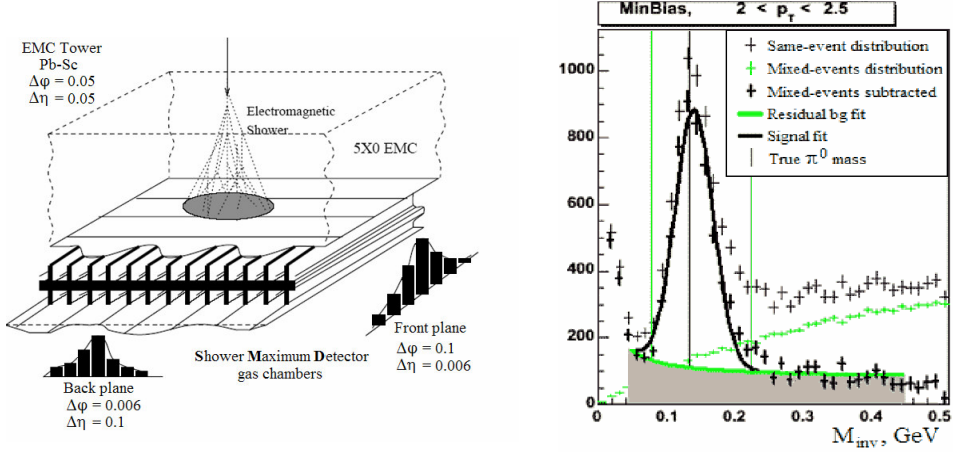

Fig. 1. (Left) Electromagnetic shower detection in the STAR calorimeter. (Right) Invariant mass distribution of photon-like cluster pairs in the BEMC. The upper, middle and lower histograms show same-event distribution, mixed-event distribution and the difference between the two, respectively. A parametrization of the residual background is indicated by the shaded area. The full line shows a Gaussian fit to the background subtracted $\pi^0$ peak.

The combinatorial background was estimated using the event mixing technique with mixing classes based on the trigger type, the vertex $z$-coordinate and the calorimeter hit multiplicity. Above 1 GeV/$c^2$, the mixed event background describes the data well. The excess seen in the region closely above the $\pi^0$ peak (Fig. 1) is probably due to the incorrectly measured opening angle of the $\pi^0$ decays and partly due to the $\eta$ decays. The peak observed at $m_{inv} < 0.05$ GeV/$c^2$ stems from cluster splitting in the EMC towers. The mixed event background was subtracted from the signal and the residual background was parameterized by an exponential function. The yields per event were obtained by integrating the background subtracted mass peak in $p_T$ bins of 0.5 (1) GeV/c for $p_T$ below (above) 7 GeV/c. The high tower trigger spectra are normalized using the pre-scale factors and triggering rates recorded during the experiment.

Corrections for reconstruction losses and detector efficiencies were calculated using Monte-Carlo simulated neutral pions embedded in real d+Au events and weighted according to the steeply falling $p_T$ spectrum as calculated from NLO pQCD. Corrections for the trigger acceptance (few percent) were not applied. The contributions from weak decays of the $K^0$s are not corrected for but estimated to be small. The overall systematic uncertainties due to the efficiency determination, yield extraction, normalization procedure and energy calibration are estimated to be 30% (50%) for $p_T$ below (above) 9 GeV/c.

The inclusive $p_T$ distribution of neutral pions is shown in Fig. 2. The yields from the different trigger samples agree within errors in the $p_T$ regions where they overlap. The obtained $p_T$ spectrum is compared with previous STAR measurements on charged pion production [9]. Except for the lowest $p_T$ point, these data agree within errors.

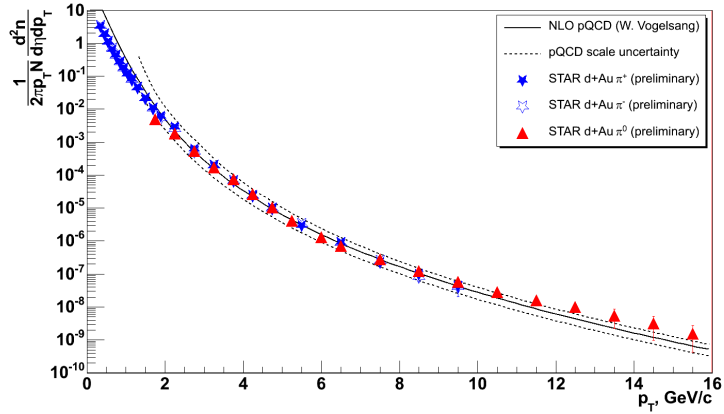

Fig. 2. $\pi^0$ inclusive $p_T$ spectrum in d+Au collisions at $\sqrt{s_{NN}}$ = 200 GeV/c measured by STAR. Also shown are charged pion results from STAR (preliminary). The full curve corresponds to the pQCD prediction described in the text. The band indicates the scale uncertainties.

The results are also compared to NLO pQCD calculations [10] using the CTEQ6M [11] parton distributions for deuterium and the nuclear parton distributions from [12] for Au. The factorization scale was set equal to $p_T$ and varied by a factor two to estimate the scale uncertainties. The fragmentation functions are taken from [13]. The Cronin effect is not included in the calculations. It is seen from Fig. 2 that the measurements are consistent with the pQCD prediction up to $p_T$ = 16 GeV/c.